\documentclass[12pt]{article}
\usepackage{authblk}
\usepackage{amsmath}

\usepackage{graphicx}% Include figure files
\usepackage{dcolumn}% Align table columns on decimal point
\usepackage{bm}% bold math
\usepackage[natbib=true,style=numeric,sorting=none]{biblatex}
\usepackage[right=2.5cm, bottom=2.5cm, left=2.5cm, top=2cm]{geometry}

\usepackage{xcolor}

\begin{document}

\title{Silicon Photomultipliers for Detection of Photon Bunching Signatures}% Force line breaks with \\

\author[1, 2]{\small Lucas Finazzi\footnote{Lead author: lfinazzi@unsam.edu.ar}}
\author[1]{Federico Izraelevitch\footnote{Corresponding author: fhi@unsam.edu.ar}}
\author[1]{Alexis Luszczak}
\author[2]{Thomas Huber}
\author[2]{Andreas Haungs}
\author[1]{Federico Golmar}

\affil[1]{Institute for Physical Sciences, University of San Martin, CONICET, Buenos Aires, Argentina}
\affil[2]{Institute for Astroparticle Physics, Karlsruhe Institute of Technology, Germany}

\date{\today}

\maketitle

%%%%%%%%%%%%%%%%%%%%%%%%%%%%%
%%%%%%%%%%%%%%%%%%%%%%%%%%%%%

\begin{abstract}

In this work, photon bunching from LED light was observed for the first time using SiPMs. The bunching signature was observed with a significance of $7.3~\sigma$ using 97~hs of data. The light was spectrally filtered using a 1~nm bandpass filter and an Etalon filter to ensure temporal coherence of the field and its coherence time was measured to be $\tau_C = (19 \pm 2)$~ps. The impact of SiPM non-idealities in these kinds of measurements is explored, and we describe the methodology to process SiPM analog waveforms and the event selection used to mitigate these non-idealities. 

\end{abstract}

%%%%%%%%%%%%%%%%%%%%%%%
%%%%%%%%%%%%%%%%%%%%%%%
\section{Introduction} 

From biomolecules to astronomical bodies, from LIDAR to Deep Space Optical communications, photon detection is a tool with wide range of scientific and technological applications. Thus, the development of new sensors enables new discoveries and pushes the frontier of knowledge in several disciplines. The ultimate photo-detector should be capable of resolving single photons with a high dynamic range, high detection efficiency and short response time, have photon-number resolution and operate at room temperature. In recent years, a novel device called Silicon Photomultiplier (SiPM)~\cite{nepomuk_3_sipms, sipm_review} was developed that, in principle, fulfills all these requirements. Currently, SiPMs have been successfully implemented in medical imaging~\cite{pet1, pet2}, particle physics detectors~\cite{hep1, hep2, hep3, hep4}, astrophysics~\cite{ap1, ap2} and communications applications~\cite{comm1, comm2}, among others. Nevertheless, applications to Quantum Optics and Photonics have been hindered in the past due to some of the SiPM's non-idealities, which can be classified in two: dark counts and correlated noise. Dark counts are generated due to thermal excitation of electrons in the Silicon and so they are generated even if no light is incident on the sensors. Correlated noise has two distinct sources: crosstalk and afterpulsing. Crosstalk events occur when a primary dark or photon count produces a prompt secondary avalanche in the Silicon. Such events have an amplitude of two detected photons. In turn, afterpulsing are events in which, after a primary dark or photon count, a delayed avalanche is generated, producing two distinct events. These effects are called correlated noise because they require a primary count to appear and they degrade the sensor performance. Recently, the impact of these features in the field of quantum optics has been studied~\cite{chesi2019}.

One of the landmark experiments that gave rise to the field of quantum optics was the measurement of intensity fluctuations of light detected from thermal sources, first done by Hanbury-Brown and Twiss~\cite{hbt_1956}. They observed photon correlations by measuring the arrival time difference between them at two separate detectors. These correlations are referred to as HBT effect or photon bunching. The name ``photon bunching'' alludes to the fact that a fraction of the detected photon events have an increased probability of arriving together. Besides opening a new field of research and giving an insight into the nature of light, Hanbury-Brown and Twiss applied this technique to astronomy and used it to measure the angular diameters of several stars~\cite{hbt}.

Nowadays, photon correlation measurements are applied to several fields, like time-correlated photon detection in time-domain diffuse optics~\cite{dalla_mora}, quantum version of Time-of-Flight LIDAR~\cite{tan_2023} and quantum image scanning microscopy~\cite{lubin_2019}. In addition, advances in instrumentation technology have given new life to the application of the HBT technique in astrophysics. For example, the VERITAS~\cite{abeysekara_2020} and MAGIC~\cite{delgado2021} collaborations have measured the angular diameter of various stars, and prospects for Cherenkov Telescope Array (CTA) were studied~\cite{dravins2012}. Moreover, the use of the technique was recently proposed for measuring gravitational waves~\cite{park2021}.

In this work we demonstrate a practical use of SiPMs for Quantum Optics. We measured photon bunching from a thermal source with these novel sensors for the first time.

%%%%%%%%%%%%%%%%%%%%%%%%%%%%%
%%%%%%%%%%%%%%%%%%%%%%%%%%%%%
\section{Signal-to-Noise Ratio}

The usual sensors used in Quantum Optics are the avalanche photodiodes (APDs). The traditional way to measure photon bunching is to use a beamsplitter and two sensors that are triggered in coincidence~\cite{tan_paper, tan_thesis} and then measure the time difference between photon arrivals, $dt$. In this approach, a Signal-to-Noise Ratio (SNR) can be defined as the ratio between bunched events to total coincidence fluctuations in a resolving time $\tau_R$. The resolving time is the overall time resolution of the experimental setup. The main contribution to this time is the 2-SiPM's composite jitter. The total coincidence fluctuations are just the Poisson fluctuations of total counts (and so they can be calculated as $\sqrt{\mathrm{total \ counts}}$). These have two contributions: The accidental (non-bunched photons plus dark counts) event fluctuations and the signal (bunched photons) event fluctuations. It is usually the case that the accidental count fluctuations are much larger than the signal fluctuations, so these are often not considered in this calculation. With this in mind, the SNR can be calculated as~\cite{hbt}

\begin{equation}
    \mathrm{SNR}_{\mathrm{APD}} = \tau_c \sqrt{R_1 R_2} V^2 \sqrt{\frac{T}{\tau_R}} \ ,
    \label{eq:snr}
\end{equation}

\noindent where $\tau_c$ is the coherence time of the field, $R_i$ is the photon rate in detector $i$, $V$ is the degree of spatial coherence, $T$ is the integration time and $\tau_R$ is the resolving time of the experiment. The coherence time is the time in which the phase of the field remains correlated or predictable, and the integration time is the total measurement time of the experiment. This SNR formula is valid for linearly polarized light, which will be the case for our experiment.

A time-difference histogram with the registered events will have a background of accidental events and a bunching excess peak, centered at zero time-difference if the optical path of both sensors is the same. When normalized using the accidental background~\cite{mandel}, this histogram is the second order correlation function $g^{(2)}$. As previously mentioned, the background fluctuations are accidental (random) coincidences of sensor counts, which come from non-correlated light and dark counts. In a case where the sensor has non-negligible dark counts, like in an SiPM, these have to be taken into account. If the resolving time $\tau_R$ is much larger than the coherence time $\tau_c$, the rate of coincidences $R_c$ in a window of duration $\tau_R$ is~\cite{scarl_contrast}

\begin{equation}
    \begin{aligned}
        R_c &= \overbrace{N_1 N_2 \tau_R }^\text{accidental} + \overbrace{R_1 R_2\tau_c V^2}^\text{signal} \\
        &= N_1 N_2 \tau_R \Big( 1 + \frac{R_1 R_2}{N_1 N_2}\frac{\tau_c}{\tau_R} V^2 \Big) \ ,
    \end{aligned}
    \label{eq:signal_accidental}
\end{equation}

\noindent where $D_i$ is the Dark Count Rate (DCR) in detector $i$ and $N_i = R_i + D_i$ is the total rate in detector $i$. Note that the presence of bunching results in an excess of events over the accidental background $N_1 N_2 \tau_R$. From Equation~(\ref{eq:signal_accidental}), it is possible to obtain a relation between the SNR using APDs with respect to using SiPMs as

\begin{equation}
    \mathrm{SNR}_{\mathrm{SiPM}} = \frac{R_1 R_2 \tau_c V^2 T}{\sqrt{N_1 N_2 \tau_R T}} = \frac{1}{\sqrt{F}} \mathrm{SNR}_{\mathrm{APD}} \ .
    \label{eq:snr_sipm}
\end{equation}

\noindent Above we defined the factor $F = \frac{N_1 N_2}{R_1 R_2}$, and we assumed negligible dark counts for the APD case. To derive the above expression, the signal counts were divided by the fluctuation of accidental counts. Equation~(\ref{eq:snr_sipm}) shows the impact of this non-ideality of SiPMs sensors in these kinds of measurements. In addition, the ratio of bunching events to accidental events in a time $\tau_R$~\cite{mandel} will also decrease when comparing the APD and the SiPM cases:

\begin{equation}
    C_{\mathrm{SiPM}} = \frac{R_1 R_2}{N_1 N_2} \frac{\tau_c}{\tau_R} V^2 = \frac{1}{F} C_{\mathrm{APD}} \ \ .
    \label{eq:height}
\end{equation}

It can be seen that $C_{\mathrm{SiPM}}$ and $\mathrm{SNR}_{\mathrm{SiPM}}$ approach the APD case when $F \sim 1$ (or $R_i \gg D_i$).

%%%%%%%%%%%%%%%%%%%%%%%%%%%%%
%%%%%%%%%%%%%%%%%%%%%%%%%%%%%
\section{Experimental Setup} 

\begin{figure}[h!]
    \begin{center}
        \includegraphics[width=10cm]{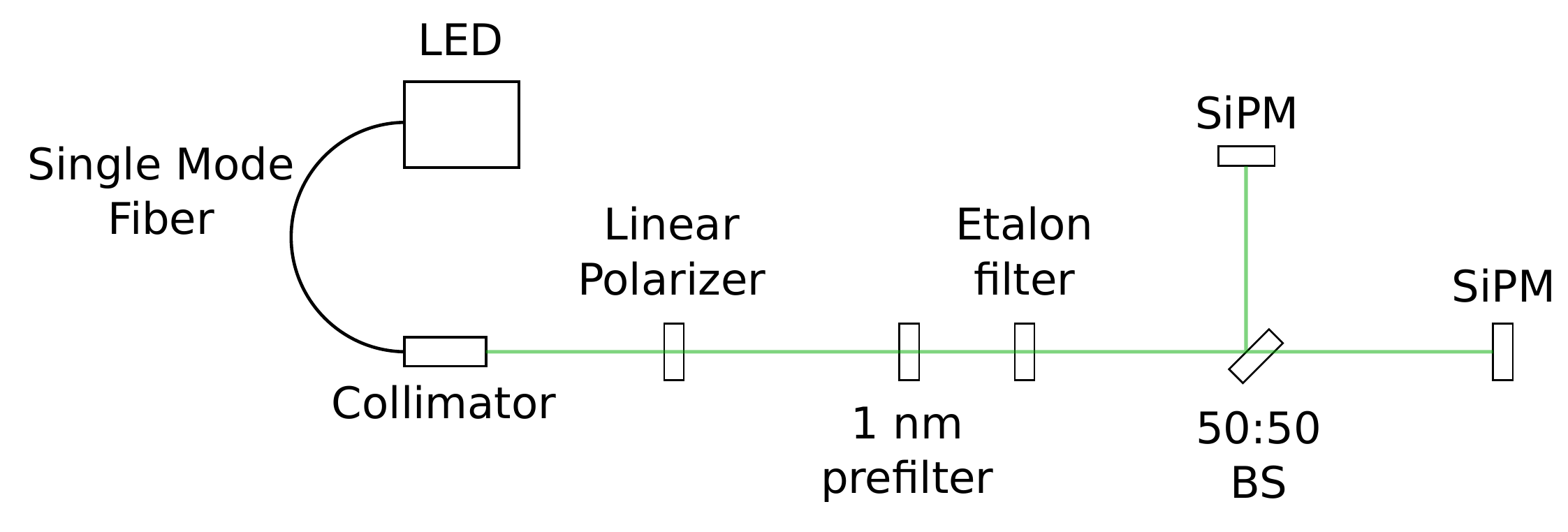}
        \caption{Schematic of the experimental setup used to detect photon bunching with SiPMs. A green LED source was coupled to a single-mode fiber to ensure spatial coherence of the field. Light was filtered with a 1~nm filter and an Etalon filter to ensure temporal coherence as well. Two 1~mm$^2$ SiPMs were placed in both arms of a 50:50 beamsplitter to perform coincidence measurements.}
        \label{fig:exp_setup_schematic}
    \end{center}
\end{figure}

The experimental setup used to measure photon bunching using SiPMs is shown in Figure~\ref{fig:exp_setup_schematic}. The setup was mounted inside a light tight facility specifically designed for SiPM characterization studies~\cite{spock1, spock2}. 

The light source used for the experiment was a fiber coupled LED from THOR LABS (TL M530F2) with a center wavelength of 530~nm. Its output was fed into a single mode optical fiber, which is used to enforce spatial coherence of the field. A rate of $10^{10}$~photons per second was obtained at the output of the fiber collimator. A linear polarizer was placed in the beam path. Then, to increase the coherence time of the field, light was spectrally filtered using a 1~nm FWHM filter (TL FL05532-1) and an etalon filter (Light Machinery, OP-7423-1686-1). The etalon free spectral range is such that approximately 17 peaks fall in the 1~nm FWHM optical passband of the interference filter~\cite{etalon_spectrum}. The coherence time of this configuration was estimated to be of the order of 10~ps~\cite{mandel}. Finally, the beam was separated with 50:50 beamsplitter and both SiPMs were placed at the same distance from it. Each SiPM (ONSEMI MicroFC-10035-SMT) was connected to an Analog Front-End board designed in-house with a time resolution below~1~ns. SiPMs were biased at 5~V overvoltage using a SourceMeter Unit and the SiPM DC current was monitored during the whole experiment. No temperature control on them was performed, but the experiment room was kept at $(22 \pm 1)$~$^\mathrm{o}$C. SiPMs were both placed at a distance of $(85 \pm 1)$~cm from the collimator output. SiPM analog waveforms were acquired using a digitizer at 500~MSps (PicoScope 2406B). The DAQ is configured with a coincidence window of 100~ns, a record length of 200~ns and a pre trigger region of 95~ns. This means that at least one of the pulses in a captured event is approximately centered in the acquisition window. A 3~ns cable delay was placed in one of the SiPM channels to help rule out hardware glitches.

Two experiments were performed: First, a signal measurement where the field was both spatially and temporally coherent (and thus bunching could be observed). Then, a background measurement where the effect was not present, because the field was purposely not temporally coherent. To remove the temporal coherence of the field, the 1~nm filter and the Etalon filter were removed and replaced with an optical attenuator to avoid SiPM signal saturation. In the case where photon bunching is not present, the background is expected to be uniform at small time differences.

The DCR and the detected photon rate on each SiPM for both experiments is shown in Table~\ref{tab:sipm_rates}.

\begin{table}[!h]
    \centering
        \begin{tabular}{ | c | c | c | c | }
        \hline
        SiPM & DCR [kcps] & \begin{tabular}{@{}c@{}}Signal experiment \\ rate [kcps]\end{tabular} & \begin{tabular}{@{}c@{}}Bkgnd. experiment \\ rate [kcps] \end{tabular} \\ 
        \hline
        1 & 126 $\pm$ 1 & 629 $\pm$ 1 & 517 $\pm$ 1\\ 
        2 & 148 $\pm$ 1 & 446 $\pm$ 1 & 522 $\pm$ 1\\ 
        \hline
        \end{tabular}
    \caption{SiPM parameters and rates for signal and background experiments.}
    \label{tab:sipm_rates}
\end{table}

For reference, these values would result in $F \sim 1.6$. Compared to the use of APDs, this results in an SNR and a bunching-peak relative height 20~\% and 38~\% lower, respectively.

The measured coincidence rate between the two detectors in the 100~ns coincidence gate was about 10~kcps. Considering the acquisition dead time, both signal and background experiments were run for an effective time of 97~hs. Using Equation~(\ref{eq:snr_sipm}), the expected SNR for that effective acquisition time is 80, while it would be 100 in the APD case.

%%%%%%%%%%%%%%%%%%%%%%%%%%%%%
%%%%%%%%%%%%%%%%%%%%%%%%%%%%%
\section{Event Selection}

\begin{figure}[!ht]
    \begin{center}
        \includegraphics[width=10cm]{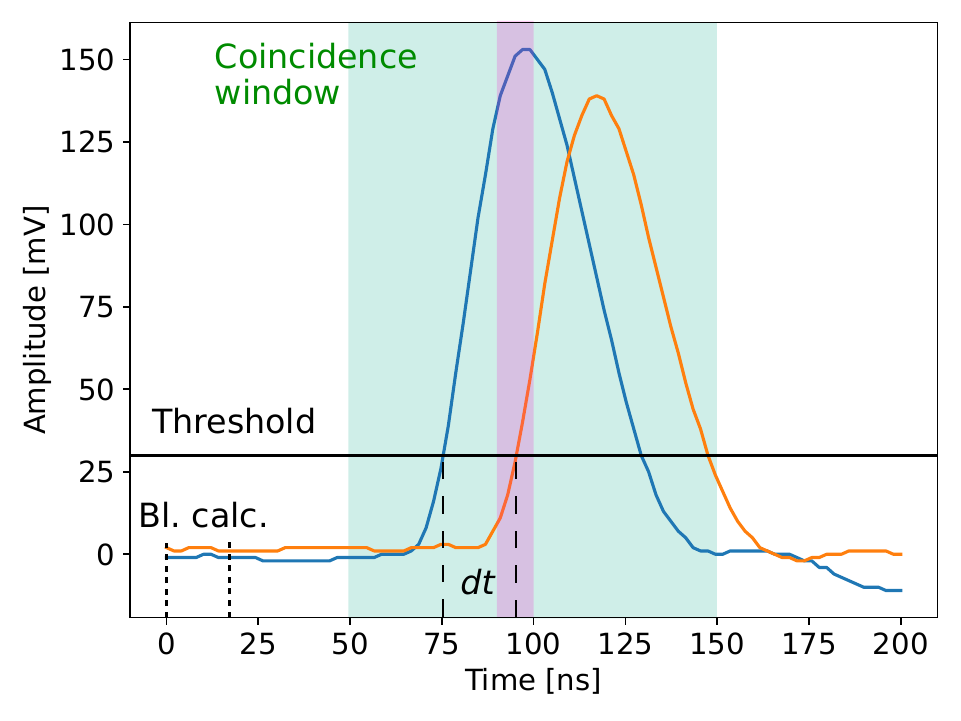}
        \caption{Example of an acquired event. The coincidence window of 100~ns is shown in green and the threshold for Timestamp determination is shown in black. The time difference between events, $dt$, in both SiPMs is shown as well. The Baseline level (Bl) was determined using the first 16~ns of each waveform. An additional window of 10~ns width and centered in 95~ns, which was used for posterior data selection, is shown in violet.}
        \label{fig:wf_cuts}
    \end{center}
\end{figure}

For each event, several parameters were calculated from the digitized waveforms acquired. These parameters were: the Baseline average and its standard deviation, Trigger Timestamp, Time-over-Threshold (ToT), waveform Amplitude, and waveform Maximum and Minimum values. These parameters were used for posterior event selection. An example of an acquired event can be seen in Figure~\ref{fig:wf_cuts}.

All parameters were calculated using digital pulse processing. To determine the Baseline mean and its standard deviation, the first 16~ns of each event were used. The Timestamp of the detected photon was determined using a linear interpolation of the waveform and a software leading-edge discriminator. The Threshold used for the discriminator was set 30~mV over the previously calculated Baseline level. To calculate the signal ToT, a falling-edge discriminator was used to obtain the second intercept between the waveform and the Threshold. The ToT was obtained as the difference between the Trigger Timestamp and the falling edge intercept. The waveform Amplitude was calculated as the waveform Maximum minus the Baseline level.

Only events with the following properties were accepted to be included in the time-difference histogram:

\begin{enumerate}
    \item Baseline standard deviation lower than 4~mV. This selection cut removes events with noisy or fluctuating baselines. An unstable baseline degrades the time resolution of the Trigger Timestamp determination. 
    \item At least one of the Trigger Timestamps present inside a 10~ns window centered at 95~ns. Due to the way the digitizer acquisition was configured, at least one of the channel triggers is expected inside this window. This selection window is shown in violet in Figure~\ref{fig:wf_cuts}.
    \item No samples above the threshold on the baseline calculating region. This selection cut is employed because the baseline can't be correctly estimated if a previous pulse is present in that region.
    \item Events with $\mathrm{ToT} < 80$~ns. Large values of ToT indicate a spurious pile-up event and were rejected. 
    \item Waveforms with an amplitude less than 200~mV. This selects events with only one detected photon -or dark count-. This removes events with crosstalk or events with two coincident photon detections on the same SiPM.
\end{enumerate}

%%%%%%%%%%%%%%%%%%%%%%
%%%%%%%%%%%%%%%%%%%%%%
\section{Results and Discussion}

\begin{figure}[!ht]
    \begin{center}
        \includegraphics[width=10cm]{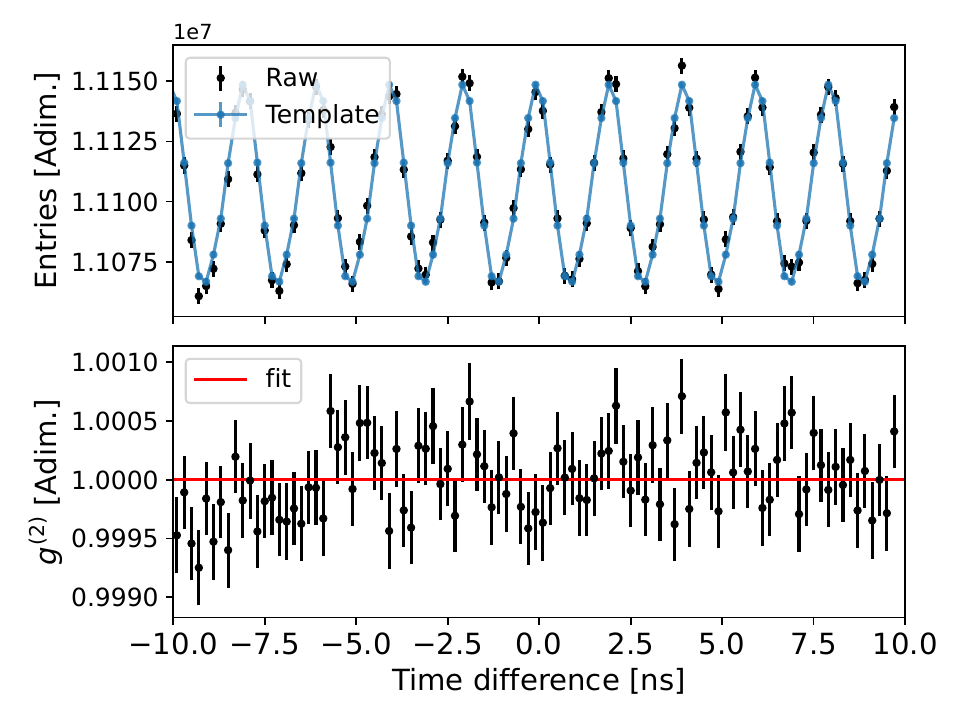}
        \caption{Top plot: raw data histogram of timestamp differences in the background experiment, along with the generated template used for the removal of the systematic 2-ns structure. Bottom plot: normalized timestamp difference histogram in the background experiment and a uniform fit with resulting $\chi^2/ \mathrm{dof} = 99.7/100$ and a p-value of 0.49.}
        \label{fig:2ns_structure}
    \end{center}
\end{figure}

After event selection, the time-difference histogram was constructed for the two experiments performed. On both histograms, an unwanted sinusoidal-like systematic pattern with a period of 2~ns was observed. The reason for this effect is that the digitizer-board channels are only synchronized down to its sampling period of 2~ns. The top histogram of Figure~\ref{fig:2ns_structure} shows the raw data points with this structure. The amplitude of this electronic jitter is negligible compared to the mean entry number, but relevant for observing the bunching peak. Due to this, a procedure had to be devised to remove this systematic effect. First, the time-difference histogram of the background dataset was cut into 2-ns slices. Then, all these slices were averaged into a single 2-ns window. Only systematic artifacts survive this procedure, so this is a good method to estimate this structure generated from the imperfect DAQ hardware. This averaged 2-ns window was then repeated along the time-difference axis to build a template of the systematic structure. This template was then used to normalize the raw histogram, and it is shown on the top plot of Figure~\ref{fig:2ns_structure}. On the bottom plot of that same figure, the normalized histogram is shown. A uniform fit was performed on the normalized background with a resulting $\chi^2/ \mathrm{dof} = 99.7/100$ and a p-value of 0.49. In addition, a Kolmogorov-Smirnoff (KS) test~\cite{ks_test} was run to compare the background to a uniform distribution, and a p-value of 0.37 was obtained. Both tests are consistent with the expected background distribution in the absence of bunching signatures. 

\begin{figure}[!ht]
    \begin{center}
        \includegraphics[width=10cm]{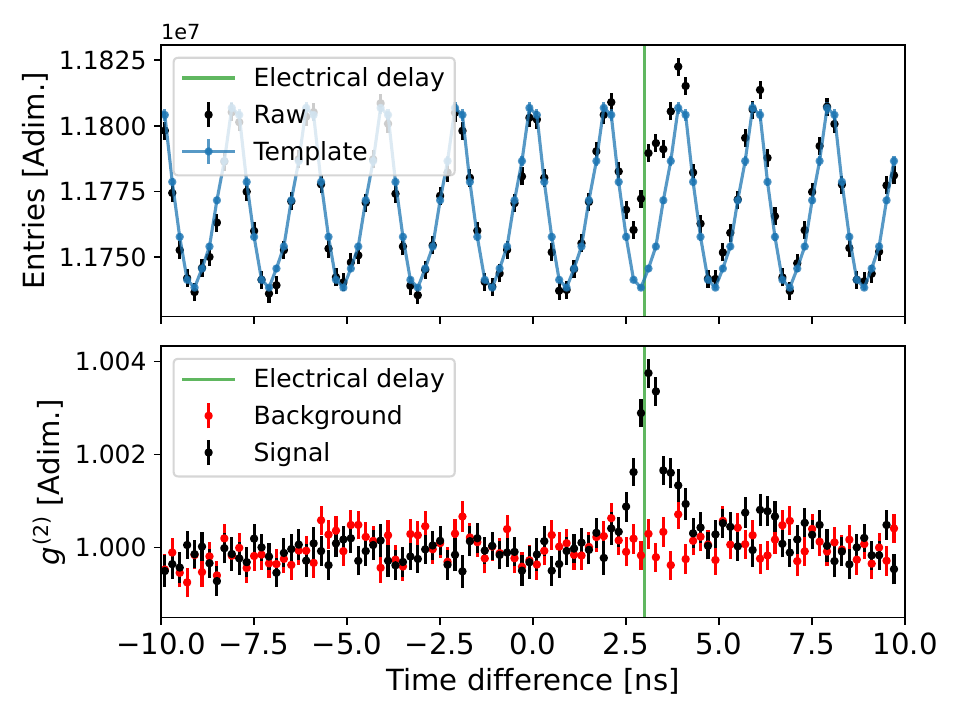}
        \caption{Top plot: raw data histogram of timestamp differences in the signal experiment, along with the generated template used for the removal of the systematic 2-ns structure. Bottom plot: normalized timestamp difference histogram of both experiments. The electrical delay of 3~ns can be seen in green.}
        \label{fig:2ns_structure_signal}
    \end{center}
\end{figure}

In the signal experiment, a bunching peak is expected to appear around 3~ns, as the optical path of both sensors was the same and a cable delay was introduced in one of the channels. The top plot of Figure~\ref{fig:2ns_structure_signal} shows the raw time-difference histogram for the signal experiment and the constructed template for the removal of the 2-ns structure. An excess of events over the template can be seen around 3~ns. As in the background case, the template was constructed slicing the raw histogram into 2-ns windows. However, two windows where the bunching peak was expected were omitted for its construction. The bottom plot of Figure~\ref{fig:2ns_structure_signal} shows the normalized histograms for both experiments. The bunching signature can be seen over the uniform background. 

Again, a KS test was performed against a uniform distribution on the signal time-difference histogram. The probability that the signal peak observed was caused by a background fluctuation is $10^{-14}$, which results in a significance level of $7.3~\sigma$. 

\begin{figure}[!h]
    \begin{center}
        \includegraphics[width=10cm]{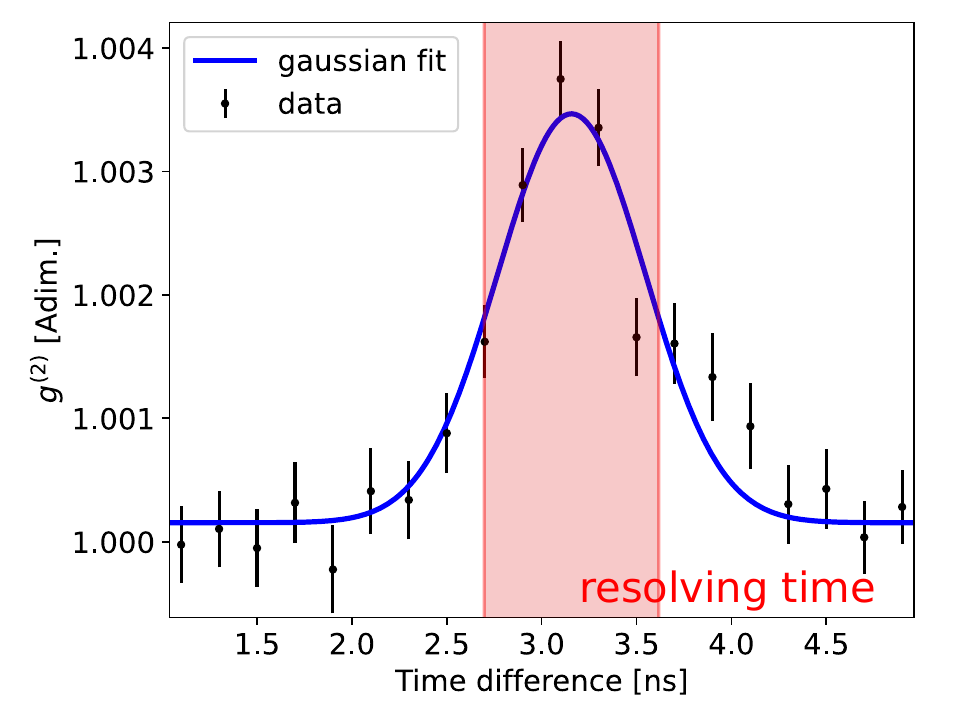}
        \caption{Signal peak with a Gaussian fit. The peak is expected to be Gaussian because it is dominated by the overall time resolution of the detection system. The fit has a $\chi^2/\mathrm{dof} = 18/16$, which results in a p-value of 0.31. The width of the peak is $(920 \pm 83)$~ps FWHM, the peak's center is $(3.15 \pm 0.03)$~ns and the peak height is $(3.3 \pm 0.2) \cdot 10^{-3}$. The system resolving time, given by the composite jitter of the two SiPM detectors is shown in red (centered around the peak maximum).}
        \label{fig:gaus_fit}
    \end{center}
\end{figure}

To further identify the characteristics of the peak, a Gaussian fit was performed on it. The coherence time of the field is much smaller that the overall time resolution of the detection system, which is a combination of SiPM and electronic jitter and the digital pulse-processing algorithms. This is the reason why the shape of the bunching peak is expected to have a Gaussian distribution~\cite{sipm_jitter_1}. Figure~\ref{fig:gaus_fit} is a zoom of the time-difference histogram centered in the bunching peak, where a Gaussian fit was performed ($\chi^2/\mathrm{dof} = 18/16$ and p-value of 0.31). 

The width of the Gaussian distribution obtained from the fit is $(920 \pm 83)$~ps FWHM, which represents the overall time resolution of the system, or the resolving time $\tau_R$. The mean obtained from the fit is $(3.15 \pm 0.03)$~ns, compatible with the $\sim 3$~ns cable delay introduced in one of the channels. The height obtained from the fit is $(3.3 \pm 0.2) \cdot 10^{-3}$. From the latter, and using Equation~(\ref{eq:height}), the coherence time $\tau_c = (19 \pm 2)$~ps was obtained. This value is consistent with the expected value in the order of 10~ps. The experimental SNR was calculated to be 99 using events in one resolving time. This value is of the same order of magnitude as the preliminary estimation given with Equation~(\ref{eq:snr_sipm}). 

%%%%%%%%%%%%%%%%%%%%%%
%%%%%%%%%%%%%%%%%%%%%%
\section{Conclusions and outlook}

In this work, photon bunching from LED light was measured using SiPM sensors for the first time with a 7.3~$\sigma$ significance and an SNR of 99. The coherence time of the field was measured to be $\tau_c = (19 \pm 2)$~ps. We presented the impact of the non-idealities of SiPMs in the equations that model this phenomenon, compared to the traditional APD sensors. In SiPMs, Dark Count Rate halves approximately every $8$~$^\mathrm{o}$C to $10$~$^\mathrm{o}$C, depending on the particular model. If they were operated at $-40$~$^\mathrm{o}$C, dark count rate would be in the order of 1~kHz, which is comparable with APD detectors. Cooled-down operation of SiPMs would enable their use for quantum optics experiments with dimmer sources. Furthermore, we described the methodology to process the analog waveform of the SiPMs, and the event selection used that mitigate the non-idealities of these sensors. 

The research presented here involved an analysis of the events with an amplitude of 1 photoelectron, i.e. one detected photon or a dark count per SiPM channel. In further works, the use of the photon-number resolution of the SiPM sensor will be explored. Particularly, the impact of the correlated noise will be studied in regards to photon bunching. This analysis could open a door for observation of higher order correlations using only 2 detectors.

\section*{Acknowledgements}

The authors acknowledge financial support from ANPCyT, PICT 2017-0984 ``Componentes Electrónicos para Aplicaciones Satelitales (CEpAS)'', PICT-2018-0365 ``LabOSat: Plataforma de caracterización de dispositivos electrónicos en ambientes hostiles'', PICT-2019-2019-02993 ``LabOSat: desarrollo de un Instrumento detector de fotones individuales para aplicaciones espaciales'', UNSAM-ECyT FP-001 and the Helmholtz research programs MU and MT at KIT. This work was done in the frame of and supported by the Strategic Partnership UNSAM-KIT (SPUK).

\printbibliography

@book{mandel,
    author = "L. Mandel and E. Wolf",
    title = "Optical Coherence and Quantum Optics",
    publisher = "Cambridge University Press",
    year = "1995" 
}

@book{hbt,
    author = "R. {Hanbury Brown}",
    title = "The Intensity Interferometer",
    publisher = "Halsted Press",
    year = "1974" 
}

@book{etalon_spectrum,
    author = "L. A. Coldren and S. W. Corzine and M. L. Mašanović",
    title = "Diode Lasers and Photonic Integrated Circuits",
    publisher = "John Wiley and Sons, Inc.",
    year = "2012" 
}

@phdthesis{tan_thesis,
    author = "P. K. Tan",
    title = "Stellar Temporal Intensity Interferometry",
    school = "National University of Singapore",
    year = "2015",
    type = "{Ph.D.} thesis", 
}

@article{tan_paper,
    author = "P. K. Tan and G. H. Yeo and H. S. Poh and A. H. Chan and C. Kurtsiefer",
    title = "Measuting Temporal Photon Bunching in Blackbody radiation",
    journal = "The Astrophysical Journal Letters",
    volume = "789",
    pages = "L10",
    year = "2014"
}

@article{scarl_contrast,
  title = "Measurements of Photon Correlations in Partially Coherent Light",
  author = "D. B. Scarl",
  journal = "Phys. Rev.",
  volume = "175",
  issue = "5",
  pages = "1661--1668",
  year = "1968"
}

@article{nepomuk_3_sipms,    
    title = "Characterization of Three High Efficiency and Blue Sensitive Silicon Photomultipliers",
    author = "A. Nepomuk Otte and D. Garcia and T. Nguyen and D. Purushotham",
    journal = "Nuclear Instruments and Methods in Physics Research Section A: Accelerators, Spectrometers, Detectors and Associated Equipment",
    volume = "846",
    pages = "106-125",
    year = "2017",
}

@book{ks_test,
    author = "W. W. Daniel",
    title = "Applied Nonparametric Statistics",
    publisher = "PWS-KENT Pub.",
    year = "1990"
}

@article{sipm_jitter_1,
    title = "Single photon time resolution of state of the art {SiPMs}",
    author = "M. V. Nemallapudi and S. Gundacker and P. Lecoq and E. Auffray",
    journal = "Journal of Instrumentation",
    volume = "11",
    pages = "P10016",
    number = "10",
    year = "2016"
}

@article{dalla_mora,
    title = "The {SiPM} revolution in time-domain diffuse optics",
    author = "A. {Dalla Mora} and L. {Di Sieno} and A. Behera and P. Taroni and D. Contini and A. Torricelli and A. Pifferi",
    journal = "Nuclear Instruments and Methods in Physics Research Section A: Accelerators, Spectrometers, Detectors and Associated Equipment",
    volume = "978",
    pages = "164411",
    year = "2020"
}

@article{tan_2023,
    title = {Practical Range Sensing with Thermal Light},
    author = {P. K. Tan and X. J. Yeo and A. Z. W. Leow and L. Shen and C. Kurtsiefer},
    journal = {Phys. Rev. Appl.},
    volume = {20},
    issue = {1},
    pages = {014060},
    year = {2023}
}

@article{lubin_2019,
    title = {Quantum correlation measurement with single photon avalanche diode arrays},
    author = {G. Lubin and R. Tenne and I. M. Antolovic and E. Charbon and C. Bruschini and D. Oron},
    journal = {Opt. Express},
    volume = {27},
    number = {23},
    pages = {32863--32882},
    year = {2019}
}

@article{abeysekara_2020,
    author = {A. U. Abeysekara and W. Benbow and A. Brill et al.},
    title = {Demonstration of stellar intensity interferometry with the four {VERITAS} telescopes},
    journal = {Nature Astronomy},
    volume = {4},
    pages = {1164–1169},
    year = {2020}
}

@article{hbt_1956,
    author = {R. {Hanbury Brown} and R. Twiss},
    title = {Correlation between Photons in two Coherent Beams of Light},
    journal = {Nature},
    volume = {177},
    pages = {27-29},
    year = {1956}
}

@inproceedings{sipm_review,
  title = {SiPM's a very brief review},
  author = {Otte, A Nepomuk},
  booktitle = {International Conference on New Photo-detectors},
  volume = {252},
  pages = {001},
  year = {2016},
  organization = {SISSA Medialab}
}

@article{pet1,
    author = {N. Otte and B. Dolgoshein and J. Hose and S. Klemin and E. Lorenz and R. Mirzoyan and E. Popova and M. Teshima},
    title = {The {SiPM} — A new Photon Detector for {PET}},
    journal = {Nuclear Physics B - Proceedings Supplements},
    volume = {150},
    pages = {417-420},
    year = {2006}
}

@article{pet2,
  author = {Gundacker, Stefan and Turtos, Rosana Martinez and Auffray, Etiennette and Paganoni, Marco and Lecoq, Paul},
  title = {High-frequency {SiPM} readout advances measured coincidence time resolution limits in {TOF}-{PET}},
  journal = {Physics in Medicine \& Biology},
  volume = {64},
  number = {5},
  pages = {055012},
  year = {2019}
}

@article{hep1,
    author = {Felix Sefkow and Frank Simon and on behalf of the CALICE collaboration}, 
    title = {A highly granular {SiPM}-on-tile calorimeter prototype},
    journal = {Journal of Physics: Conference Series},
    year = {2019},
    month = {1},
    volume = {1162},
    number = {1},
    pages = {012012},
}

@article{hep2,
  title = {Measurement of the positive muon anomalous magnetic moment to 0.46 ppm},
  author = {Abi, Babak and Albahri, T and Al-Kilani, S and Allspach, D and Alonzi, LP and Anastasi, A and Anisenkov, A and Azfar, F and Badgley, K and Bae{\ss}ler, S and others},
  journal = {Physical Review Letters},
  volume = {126},
  number = {14},
  pages = {141801},
  year = {2021}
}

@article{hep3,
  title = {A light tracker based on scintillating fibers with {SiPM} readout},
  author = {M. N. Mazziotta and C. Altomare and E. Bissaldi et al.},
  journal = {Nuclear Instruments and Methods in Physics Research Section A: Accelerators, Spectrometers, Detectors and Associated Equipment},
  volume = {1039},
  pages = {167040},
  year = {2022}
}

@article{hep4,
  title = {Design and characterization of the {SiPM} tracking system of {NEXT}-{DEMO}, a demonstrator prototype of the {NEXT}-100 experiment},
  author = {V. {\'A}lvarez and M. Ball and F. I. G. Borges et al.},
  journal = {Journal of Instrumentation},
  volume = {8},
  number = {05},
  pages = {T05002},
  year = {2013}
}

@article{ap1,
  title = {Cosmic-ray isotope measurements with {HELIX}},
  author = {Allison, P and Beatty, JJ and Beaufore, L and Chen, Y and Coutu, S and Ellingwood, E and Gebhard, M and Green, N and Hanna, D and Kunkler, B and others},
  journal = {Proceedings of Science},
  volume = {358},
  year = {2019}
}

@article{ap2,
    author = {G. Ambrosi and V. Vagelli},
    title = {Applications of silicon photomultipliers in ground-based and spaceborne high-energy astrophysics},
    journal = {The European Physical Journal Plus},
    volume = {137},
    pages = {170},
    year = {2022}
}

@article{comm1,
  title = {The future prospects for SiPM-based receivers for visible light communications},
  author = {Zhang, Long and Chun, Hyunchae and Ahmed, Zubair and Faulkner, Grahame and O'Brien, Dominic and Collins, Steve},
  journal = {Journal of Lightwave Technology},
  volume = {37},
  number = {17},
  pages = {4367--4374},
  year = {2019}
}

@article{comm2,
  title = {A {SiPM}-based {VLC} receiver for Gigabit communication using {OOK} modulation},
  author = {Ahmed, Zubair and Singh, Ravinder and Ali, Wajahat and Faulkner, Grahame and O’Brien, Dominic and Collins, Steve},
  journal = {IEEE Photonics Technology Letters},
  volume = {32},
  number = {6},
  pages = {317--320},
  year = {2020}
}

@article{chesi2019,
    author = {G. Chesi and L. Malinverno and A. Allevi and R. Santoro and M. Caccia and A. Martemiyanov and M. Bondani},
    title = {Optimizing Silicon photomultipliers for Quantum Optics},
    journal = {Scientific Reports},
    volume = {9},
    pages = {7433},
    year = {2019}
}

@inproceedings{delgado2021,
    author = "The MAGIC collaboration",
    title = "{Intensity interferometry with the {MAGIC} telescopes}",
    booktitle = "Proceedings of 37th International Cosmic Ray Conference",
    year = 2021,
    volume = "395",
    pages = "693"
}

@article{park2021,
    author = {I.H. Park and K.-Y. Choi and J. Hwang and S. Jung and D.H. Kim and M.H. Kim and C.-H. Lee and K.H. Lee and S.H. Oh and M.-G. Park and S.C. Park and A. Pozanenko and C.D. Rho and N. Vedenkin and E. Won},
    title = {Stellar interferometry for gravitational waves},
    journal = {Journal of Cosmology and Astroparticle Physics},
    volume = {2021},
    number = {11},
    pages = {008},
    year = {2021}
}

@article{spock1,
  title = {{SiECA}: {S}ilicon photomultiplier prototype for flight with {EUSO-SPB}},
  author = {Painter, William and Bertaina, Mario and Bortone, Alberto and Haungs, Andreas and Menshikov, Alexander and Renschler, Max},
  journal = {Proceedings of 35th International Cosmic Ray Conference},
  volume = {442},
  year = {2017}
}

@article{spock2,
    author = {Max Renschler and William Painter and Francesca Bisconti and Andreas Haungs and Thomas Huber and Michael Karus and Harald Schieler and Andreas Weindl},
    title = {Characterization of {H}amamatsu 64-channel {TSV} {SiPMs}},
    journal = {Nuclear Instruments and Methods in Physics Research Section A: Accelerators, Spectrometers, Detectors and Associated Equipment},
    volume = {888},
    pages = {257-267},
    year = {2018}
}

@article{dravins2012,
    title = {Stellar intensity interferometry: Prospects for sub-milliarcsecond optical imaging},
    journal = {New Astronomy Reviews},
    volume = {56},
    number = {5},
    pages = {143-167},
    year = {2012},
    doi = {https://doi.org/10.1016/j.newar.2012.06.001},
    author = {Dainis Dravins and Stephan LeBohec and Hannes Jensen and Paul D. Nuñez},
}

\end{document}